# THE CONSUMER PRODUCT SELECTION PROCESS IN AN INTERNET AGE: OBSTACLES TO MAXIMUM EFFECTIVENESS AND POLICY OPTIONS

*Mark S. Nadel*[*]

TABLE OF CONTENTS

---

[*] Attorney/Advisor, Federal Communications Commission; B.A. 1978, Amherst College; J.D. 1981, Harvard. The views expressed in this article are solely the personal views of the author. The author wishes to thank Howard Anderson, Jim Bettman, Ray Burke, Farrell Bloch, Bob Cipes, Kristin Diehl, Douglas Galbi, David Goldstone, Ward Hanson, Tom Krattenmaker, Hank Levine, Elliot Maxwell, Bruce McCullough, Jim McConnaughey, Evan Kwerel, Andy Mirsky, Carolina Nadel, Eugene Nadel, Walter Nicholson, Maureen O'Rourke, Bob Peterson, Jeff Prisbrey, Frank Robert, John Rothchild, Barry Schwartz, Charles Steinfield, Kevin Werbach, and Rolf Wigand for their helpful comments. The author welcomes comments at msnadel@acm.org.

I.  INTRODUCTION

> *Your twelve-year-old microwave oven dies. You want a high quality replacement with a revolving tray that can hold a 10" plate and fit into a space 26" wide, 14" deep, and 13" high. Unfortunately, the most recent* Consumer Reports *analysis of microwaves is one year old, and none of the dozen models reviewed strikes your fancy. Searching the Internet drowns you with choices. You wish it were possible to consult someone with detailed knowledge of ALL microwave ovens for sale, who could quickly and easily provide web addresses for the few models that best met your criteria and were rated highest by a person or institution you respected. To save hours of searching, you would even be willing to pay a small fee.*

The Internet gives online consumers unprecedented access to detailed information about products for sale anywhere in the world. This helps many buyers discover precisely what they want and where to buy, often enabling them to purchase directly from providers (e.g., travelers buying airline tickets from airlines). Many consumers, however, overwhelmed by the multitude of choices and abundance of data about those choices, still want assistance when shopping. They seek a recommendation of their best choice or a short list of the most suitable options for further consideration.

Shoppers unsure of which product best fits their needs may currently consult various types of intermediaries (retailers, *Consumer Reports*, etc.) or online "infomediaries." In the Internet Age, the most effective infomediary for this task is likely to be one that can help consumers craft search profiles (based on their individual needs and desires) and apply them to a sufficiently comprehensive database (containing detailed information about all the products in a market segment). Databases, like the real estate industry's multiple-listing services ("MLSs"), offer consumers access to a dramatically broader set of options than any traditional store or salesperson's memory could hold. They also permit shoppers to sort these options according to dozens, if not hundreds, of attributes. Computers then enable buyers to transcend the information processing capacities of humans by applying precisely designed personal profiles to those databases and generating short, accurate lists of the best options. The Internet then enables consumers to expand those lists into virtual high-tech catalogues of their choices by linking to the Internet addresses depicting the items.

By enabling such databases to be shared globally, the Internet creates an environment conducive to the emergence of *selection assistants* ("SAs") that both compile relevant product databases and help buyers design profiles for effectively searching them. With detailed descriptions of individual products already online, SAs can offer separate, "unbundled" consulting services, without the need to finance or manage costly product displays and inventory. Emerging forms of SAs create the potential for an environment approaching perfect information and perfectly competitive markets.

While the marketplace is already providing consumers with a few versions of SAs, the government should ensure that buyers have the opportunity to enjoy the full benefits of competition and innovation that this technology offers. The U.S. Commerce Department is already attempting to identify regulations that may unnecessarily hinder e-commerce firms, but policy analysts should also consider how to prevent the emergence of corporate practices likely to stifle optimal competition and innovation. To help public officials identify such potential impediments to efficient e-business, this Article provides a draft framework for understanding the dynamics of the product selection process in an Internet environment. It then identifies three categories of potential obstacles that may prevent full and effective competition. Section III examines those that may limit the *breadth* of choices consumers can consider. Section IV discusses those that may limit consumers' ability to compare *features* of attractive alternatives. Section V considers hindrances to consumer efforts to design the most effective *search profiles*.

This Article finds that many threats to a fully competitive cyberspace environment can be handled effectively by the private sector, through adjustments to business strategies. Accordingly, this mechanism should be favored unless it appears likely to be ineffective. As one example, this Article examines the complaints of eBay and the National Association of Realtors ("NAR") that competitors are free riding on the data that they have compiled and published to attract buyers to their websites. It observes that current laws provide eBay and other database compilers with many opportunities to adopt strategies for earning significant returns on their investments. Moreover, this Article explains how the "database protection" law supported by eBay, NAR, and many others might have diminished innovation and competition. The Article also identifies multiple strategies that valued evaluation services, like *Consumer Reports,* might adopt to protect themselves against free riding.

Furthermore, this Article finds that improvements in information processing technologies justify a reevaluation of regulations that limit competition out of fear that consumers may be unqualified to make good choices. Thus, as SAs permit the public to search effectively for lawyers, doctors, etc., according to the particular skills desired, a shift from licensing restrictions to voluntary certifications would seem likely to increase competition and substantially benefit consumers. This Article also observes that other potential problems seem most amenable to voluntary industry trade association action or the entry of trusted third parties — remedies which are generally preferable to government action.

This Article recognizes, however, that in many cases at least a limited government role is in the public interest. For example, where trade associations have set reasonable business standards, they may welcome government enforcement efforts that prevent "bad apples" from tarnishing the industry's reputation. When industries develop new products with confusing features and competitors cannot agree on norms allowing easy comparisons by buyers, government efforts to adopt recommended criteria for this purpose may be helpful to both buyers and sellers of the best products. Such efforts may be particularly useful in helping industries set guidelines to avoid congestion on members' website servers caused by price bots or "spiders," a problem that eBay has already taken to court.

In its role as public educator, it also seems appropriate for the government to help consumers learn to use new information search tools for shopping as well as other purposes. This Article suggests that consumers should be advised about how to protect their ability to switch from one SA to another without losing the chance to enable their new SA to provide customized selection services based on their buying history.

The government should also clarify how existing antitrust laws against tying and denials of access to an essential facility apply to database businesses. The district court in *United States v. Microsoft* has already recognized the difficulty of applying the rules against "tying" to new information technologies. This analysis may be particularly important if "network effects" lead to the virtual monopolization of detailed product databases and multi-product collaborative filtering databases. Policymakers have also recognized that existing disclosure rules deserve to be reevaluated in light of the emergence of new information technologies.

While policymakers usually wait for anticompetitive practices to arise before acting, a more proactive strategy can be socially beneficial by anticipating and deterring harmful conduct and, thus, avoiding less productive uses of resources. Many scholars have embraced this approach, recognizing that it is desirable to promote the most vigorously competitive and innovative e-commerce environment possible.

II. BACKGROUND

*A. The Internet Has Widened the Global Marketplace*

The Internet is only the newest technology to increase consumer choices by expanding the size of manufacturers' markets. Transportation technologies — including railroads, canals, and trucks — first enabled consumers to purchase from distant merchants. The postal service and photographic technologies allowed consumers to view catalogue photographs of products that were not displayed locally. More recently, broadcast and cable television have allowed buyers to view full-motion displays of distant products. Now, the Internet makes it easy to view — on demand — sophisticated product descriptions from firms throughout the world, with language translation technologies eliminating yet another barrier. Most manufacturers have concluded that they cannot afford to ignore online shoppers.

Many consumers, long comfortable with making purchase decisions based on photographs from the more than thirteen billion catalogues mailed annually, find the abundance and quality of product information available on websites to be quite compelling. Thus, increasing numbers of consumers are willing to forgo in-person attention from salespeople, socializing with friends, serendipitous adventures, and other desirable aspects of conventional shopping. Although shoppers may prefer to stay offline when seeking a sensual item, such as a new perfume, food, or a clothing texture, online shoppers are discovering that the Internet often offers an efficient way to at least narrow down their options, if not also enjoy compelling online shopping experiences.

### B. *Shopping the Internet "Without" Middlemen Can Be Frustrating*

Shoppers can currently use Internet search engines to access manufacturers directly through the World Wide Web by using the generic name of a product or some of its features. Such "disintermediated" searching, as predicted by some commentators, however, has many drawbacks when a consumer has not yet selected the precise product to buy. In particular, the lists produced by search engines are generally both under- and over-inclusive. This happens for many reasons: not all suppliers presently maintain web sites; no single search engine appears to cover even half of the "indexable web" (at least partly due to the time lag in listing them); many sites are not labeled clearly (in some cases purposefully to deceive search engines); some sites may be purposefully excluded for competitive reasons; and searches often yield so many extraneous offerings that they overwhelm the user. To paraphrase one forecaster of retailing technology, "imagine looking for a blouse through a department store carrying nothing but unsorted blouses from the basement to the top floor." The development of web page standards, like eXtensible Markup Language ("XML"), and specialization of search engines and agents should improve the ability of shoppers to retrieve web pages with the attributes they seek, but searchers must still rely on the credibility of many different suppliers and ignore most data generated by third parties.

Consumers may also use the Internet to solicit sellers directly. They might post an electronic want ad in a publication's online edition or at a forum hosting "requests for proposals" ("RFPs"). Websites, such as Autobytel.com, which allow buyers to seek the best deal for a previously-selected product, can be very useful for finding the merchant with the best price. Consumers can also use sites like ValueStar.com, or Respond.com, which replace postings with e-mails to a list of potentially relevant sellers. If those hosting RFPs remain relatively passive, however, the solicitations may attract numerous undesirable responses, particularly when the buyer seeks subjective features, like a "charming" inn or an "authentic" autographed baseball. While ValueStar's system of screening buyers may eliminate extraneous offerings, it also increases the likelihood that the buyer will miss eager and qualified potential sellers who do not believe that ValueStar's annual "certification" fee is cost effective.

Two contrasting flaws handicap both search engines and forums for posting RFPs: they omit some desired items and fail to exclude other undesirable ones. These are the same twin failings of even the best brick-and-mortar stores. Some of the latter have tried to remedy over-inclusiveness by providing consumers with "personal shoppers," but the inherent space constraints, which limit the number of displays stores can offer, are a more difficult problem. Although personal shoppers could alleviate this restriction by using their knowledge of the entire product market, most tend to confine their suggestions to the limited selections offered by their employers. Thus, shoppers seeking an optimal choice usually need to consult multiple stores, search engines, or bulletin boards, and spend valuable time on frequently fruitless forays, something few seem willing to do. The problem faced by brick-and-mortar stores is that each offering must either be included or excluded; there is no middle ground.

### C. *What Shoppers Want and the Value of Databases*

While consumers desire to identify their best option, most willingly settle for less-than-the-best choices, a practice that has been called "satisficing." Often this is due to limitations on their ability to store and process large amounts of information, their general reluctance to spend costly time and money on searches, and their desire to satisfy other goals. Thus, shoppers commonly decline to consider greater numbers of relevant options, even though it would improve the accuracy of their selection. Instead, consumers generally construct preferences and strategies heavily based on situational considerations, such as the limited effort they are willing to expend on a search. Consumers seek to maximize their satisfaction with the combination of their choice *and* their *decision process.* Shoppers then generally use a two-stage selection process. First, they use some strategy to narrow their options from all alternatives to a manageable "consideration" set; then they devote greater efforts to select from that set.

Computers now provide consumers with a much more effective tool for storing and processing product information. They reduce the effort required to consider massive data, and produce better selections, among other effects. Moreover, given the overwhelming number of choices and the diminished time consumers are willing to spend on shopping, the need for such assistance is increasing. As the CEO of Furniture.com observed, customers "don't want a football field of furniture. . . . They want a boutique filled exclusively with the stuff they love." The best system would appear to be one that permits consumers to create a search profile based on their preferences, apply it to the set of all worthwhile options, and generate a consideration set of all of their best choices.

SAs can use databases to provide this service. By permitting electronic entries to replace the need for physical inventory, databases eliminate spatial limitations. Databases make it practical for SAs to offer buyers the opportunity to consider all possible options with searches that could range across a virtually unlimited number of categories. Furthermore, the electronic entries can be carefully formatted in rows and columns to facilitate quick, easy, and accurate sorting and to generate customized consideration sets of products. Thus, databases have long been used in "non-inventory" markets, including real estate, travel, and financial securities, and increasing numbers of traditional retailers, such as Barnes and Noble, Borders, and outdoor goods merchant REI, even permit customers to search online product databases from within their stores.

Meanwhile, the proliferation of entities offering credible performance evaluations — strongly stimulated by the Internet — enables databases to store data according to those formerly-more-difficult-to-search-by variables, which economists call "experience" and "credence" qualities of goods. Experience qualities, such as how funny a film is, are those which generally require one to use a product before making an assessment. Credence qualities, such as how perceptive a doctor, lawyer, or repair shop is, are difficult to evaluate even after repeated uses.

Brand names have long enabled manufacturers to communicate estimates of the experience and credence qualities of their products. Consumers can also judge experience and credence qualities based on comments from friends and relatives. Better yet, shoppers can rely on trusted experts, such as *Consumer Reports* and any of the new Internet-based rating services, thereby diminishing consumers' need to rely on brands, which represent the judgments of inherently biased manufacturers. Although consumers appear unwilling to expend significant effort to gather this data themselves, SAs can now monitor these evaluations and store them in an accessible database format.

Currently, many consumers are willing to pay directly for buying guides like *Consumer Reports* and Michelin tour books as well as other editing services, including those provided by travel agents, interior decorators, or even summer camp consultants. Buyers recognize the value of improving the quality of their choices and reducing their search time. They also realize that informediaries generally serve the party that pays them. Two factors, however, have hindered the emergence of this SA model as a major category of profitable e-commerce entities. First, many firms have been providing forms of this service free of charge to build their business or earn advertising revenues, while other individuals and groups have been doing so for other reasons. Against this background, consumers have been less inclined to pay even small amounts for information that they have become accustomed to receiving free of charge. This should change, however, as free services become more cluttered with advertising or disappear in bankruptcy or by starting to charge for their service. SAs may also train consumers to pay small amounts for advice by analogizing the payments to the cost of photocopying the information or tipping a gofer. In March 2001, Microsoft and Travelocity announced plans to adopt this strategy of charging for these services.

Second, while consumers might still be willing to pay consultants anywhere from fifty cents to a few dollars per transaction based on the incremental benefit of high quality information — namely, better choices and decreased search time — relatively high billing costs may make it uneconomic to bill for such small amounts. This should change, however, as micro-payment technologies improve and make it practical to bill on a less than dollar-per-minute basis or to impose some other small surcharge.

### III. ENSURING ACCESS TO ALL DESIRED OPTIONS

Some online shopping assistants, including Shopper.cnet.com, have followed the brick-and-mortar store model of only presenting products from which they can profit. Others, like mySimon.com, however, have recognized the long-term value of all-inclusiveness. They realize that customers, particularly those with unusual preferences or those inclined to regret a missed opportunity, are likely to be drawn to a site that enables them to consider *all* of the most suitable options. Thus, in the fall of 1998, Barnes & Noble boasted in advertisements that its 4.6 million-plus-book database dwarfed the 3 million-book database offered by Amazon.com, and a website planned for 2001 promises to permit travelers to view many times the number of flights that they can now. A wider breadth of options could also increase a website's potential advertising revenues.

A 1999 cover story in the *Harvard Business Review* calls SAs "navigators" and advises companies to assume that role themselves, even if it requires recommending competitors' products. The article's authors advise that, to retain customers' trust and loyalty, firms must demonstrate their willingness to serve the client's best interests by navigating through all of cyberspace and beyond. The major airlines adopted this strategy when they voluntarily included the flights of their competitors in their customer reservation systems. Similarly, the retail website proposed by General Motors will include non-GM cars, and "buyer" real estate agents have realized that knowledgeable home buyers would value the chance to expand their options to include the homes of sellers not offering to pay broker commissions.

Meanwhile, SAs should not find it too difficult to be comprehensive. Those in retailing already employ buyers to survey the Internet, trade shows, publications, and competing retailers to select from the universe of offerings. The cost to be comprehensive would only be the incremental expense of carefully noting all items already discovered and reviewed by the retailer's buyer (including any rejected as inadequate for most consumers), which should generally not be terribly burdensome. SAs could also adopt the approach of EveryCD.com, which offered free CDs to customers who identified titles missing from its database. In addition, established SAs could expect that manufacturers and other sellers would eagerly contact them to promote new products.  Given the increased availability of second-hand items, there would be no need to weed out discontinued products, and databases could even include items not "officially" for sale anywhere.  If many SAs were to compete in a market, consumers might also desire a "meta" SA to help them select among SAs.

### A. Prohibiting Misleading Claims of Coverage

Consumer demand for a comprehensive database of product information would likely lead at least one SA to provide it.  A problem arises, however, if SAs can mislead consumers about the comprehensiveness of their databases.  While current advertising guidelines may discourage such exaggerations, enforcement bodies might tolerate much looser uses of phrases like "all-inclusive" and "comprehensive" as ordinary puffing, especially given the apparent impracticality of monitoring *all* relevant options in a global product market.

Ideally, trade associations would help shoppers identify truly "all-inclusive" SAs by formulating and publicizing reasonably precise and effective definitions of the multiple variations of the term.  Alternatively, one would expect that consumer advocates like *Consumer Reports* might step up to the task and designate a seal indicating that an SA offered "comprehensive coverage," relying on government assistance for preventing misleading claims concerning these terms.  Absent that, a government agency might set reasonable criteria for such a seal.  Meanwhile, comprehensive SAs could simultaneously offer shoppers filtered lists of those items that met the SAs' quality thresholds.

### B.  Imposing Access Rules on Dominant Firms

Although some describe the e-commerce environment as a "frictionless" market, other experts observe the danger of entry barriers.  In particular, the compilation of databases appears to have some network effects along with other characteristics of natural monopolies.  Still, whether SAs emerge with substantial market
power will depend on many factors with respect to both supply and demand.  On the supply side, market power is more likely to arise in market segments where maintaining an accurate database is more expensive.  On the demand side, it will depend on whether there are enough buyers to support the cost of maintaining two or more databases.  In that vein, the development of more user-friendly consumer interfaces, particularly voice recognition technologies, could dramatically expand demand.  Some market segments could be winner-take-all markets, although specialized market niches might arise and limit the market power of the dominant generalist SA.

Even if SAs were likely to enjoy substantial market power, suppliers could be left to their own devices to secure access to dominant SAs. The market pressures for full coverage, discussed above, and fear of monopoly regulation might lead SAs to include the products of all suppliers, while simultaneously offering a filtered version of "preferred" options.  On the other hand, past refusals of media owners to grant access in other information industries, for economic and other reasons, suggest that some types of access rules might be justified to regulate SAs.

In fact, the exclusionary actions of the publisher of the Official Airline Guide ("OAG") led the Federal Trade Commission ("FTC") to bring suit to prevent the OAG from unfairly excluding some airlines from the guide.  Comparable non-discriminatory "common carrier" access rules have also been imposed on others with market power including telephone networks and the postal service. More recently, the fear of exclusionary practices by popular travel websites has led government agencies to investigate, and it appears to be one of the primary reasons that British Airways created its own website.

While regulations that burden innovation by attempting to micro-manage could be counterproductive, when SAs enjoy substantial market power, clear, general, pro-competitive, "non-discrimination" access rules might be appropriate. They probably represent a more efficient alternative to private lawsuits and non-expert courtrooms for resolving access rules in this specialized area. It might also be worthwhile to begin considering how to prevent database operators with monopoly power from tying the use of their databases to other services.

### C. Combating Efforts to Deny Buyers Options

1. Enabling Consumers to Employ Effective Aggregators

The emergence of "aggregators," firms that offer consumers lists of options aggregated from multiple primary sources, would appear to be in the public interest. Yet at least some sources have attempted to deny aggregators the right to use the information the former publish publicly. For example, the National Association of Realtors ("NAR") has fought efforts by Microsoft's HomeAdvisor.com to use NAR's affiliate, Realtor.com, as a source for listings of homes for sale. In addition, eBay fought efforts by both Auctionwatch and Bidder's Edge to aggregate data about the offerings of auction sites and even obtained an injunction preventing Bidder's Edge from repeatedly querying its website for auction listings when technical solutions did not work. Some online retailers have made comparable efforts to block access by price-searching shopping bots.

Many database compilers have focused their attention on Congress, where two bills were introduced in the House in 1999 — H.R. 354 and H.R. 1858 — to protect database compilers against unauthorized copying. Those bills responded to a 1991 Supreme Court decision, which clearly stated that the copyright law does not grant a party the right to prevent others from using its collection of unoriginal factual data, simply because of the considerable "sweat" expended to compile the information. Current copyright law only protects the novel aspect of such a presentation.

H.R. 1858 would have prohibited firms from copying data from a compiler and selling virtual copies of it in competition with that compiler. This prohibition would seem to have done little harm to consumers, as the data affected is already generally available to them free of charge. Meanwhile, the bill might have stimulated additional compilations of useful databases for consumers — filled with the descriptive data discussed below — by preventing mere copiers from siphoning away the attention of buyers and the associated advertising revenues. Over the long run, this might have improved consumer welfare. H.R. 1858 also would have permitted third parties to mine existing databases to produce *substantially enhanced* databases, such as those composed by aggregators.

H.R. 354 would have provided database owners with much stronger property rights to their data. In addition to prohibiting uses by pure copiers, H.R. 354 would have prohibited many uses by true aggregators and others developing innovative, enhanced services. In addition to concerns about its constitutionality, H.R. 354 raises some major policy questions. By preventing others from using the data collected to build innovative and substantially enhanced databases to serve consumers, it would have, to some extent, hindered the flow of publicly available information about competing products, which is necessary for a free market economic system to function most effectively. In addition, granting fact collectors strong intellectual property rights could well discourage new innovations in providing access to those facts.

Scholars and policymakers recognize the advantages of limiting the intellectual property rights provided by legislation. Only when and if new technologies encourage conduct that threatens the economic viability of valued existing or potential future enterprises would there seem to be justification for modifying the associated property rights. At present, however, this does not appear to be the case with respect to factual data about products. Absent congestion, the current intellectual property laws, supplemented by legislation like H.R.1858, would seem to provide sufficient incentives for the creation and use of an optimal level of databases.

*Serving sellers.* Real estate brokers and eBay have the ability to obtain full compensation for any valuable services they provide to sellers, including charging for advising sellers on how to market their properties and for posting their information in an online database. EBay can earn its fees by providing a superior auction mechanism, in terms of rules and execution. Additional publicity via aggregators should help sellers and those earning commissions from a sale, assuming that navigators do not create congestion problems.

*Serving buyers.* To the extent that eBay wants to earn the advertising revenues from repeated buyer visits to its website, it seems fair to expect it to offer the form of displays that buyers want. If auction buyers, like travel agents, prefer one-stop aggregated lists of offerings in a category, then, absent unreasonable congestion problems, they should be permitted to obtain them. Accordingly, eBay should consider adopting an aggregator's model by including data from competing auction sites. While this would eliminate one reason for sellers to use eBay and might lead to the loss of some sales to competing auction sites, these effects did not stop eBay from introducing a recommendation service for buyers. Meanwhile, the emergence of aggregators seems to be inevitable as well as in the public interest. Both sellers and their exclusive agents should want maximum exposure for their offerings to all bona fide potential buyers and buyer agents, just as if the databases were catalogues. Against this background, it seems appropriate for policymakers to scrutinize Realtor.com's "Golden Alliance" program providing cash payments to MLSs for *exclusive* rights to post their listings, effectively prohibiting the latter from providing listing data to competing national real estate websites.

2. Licensing and Other Government Restraints on Choice

Customer choices are also limited by government policies, such as laws hindering consumers from employing unlicensed providers of a service. For example, legal secretaries or paralegals who clearly disclose their non-lawyer status, but offer to provide minimal legal assistance at a low price, are prohibited from doing so without a professional license. While there is no doubt that consumers benefit from being able to identify which service providers or products have met established quality standards, it is not clear that there is any benefit to denying them the option to assume the risk of overriding those standards. In fact, many critics claim that licensing laws are designed primarily to protect the economic interests of members of professional groups rather than the public; that unlicensed providers often provide comparable services; and that licensing is generally bad public policy. Regardless of whether any of these contentions are true, it is important to recognize that the increasing availability of versions of SAs that are easily accessible to the general public and user-friendly at least partially undermines the primary justification for most types of licensing, even if other, more limited, rationales remain.

Most licensing or other sales prohibitions appear to assume that consumers are unable to protect themselves from unintentionally hiring unqualified individuals or using harmful products, due to inadequate information, a lack of sophistication, or the presence of significant misinformation. This assumption is valid to the extent that the services offered by professionals are difficult to assess prior to purchase because their primary attributes are the "experience" or "credence" qualities discussed above, and such assessments are often made in periods of stress, or that they are complex, multi-variable decisions.

Today, however, the selection process appears to be getting much more user friendly. Professionals can and do secure certifications of many specialized skills, and there are many new technologies for monitoring quality. In addition, professionals are not shy about publicizing their expertise and laws prohibiting such advertising have been set aside. Furthermore, many local media offer consumers comparative evaluations of professionals. Against this background, consumers will increasingly be able to use SAs armed with comprehensive databases to find professionals meeting all of their requirements, as discussed below. Once access to these SAs is easy, it would seem that specialized certificates and disclosure requirements would be more beneficial to society than absolute bans on practicing without a license.

Short of eliminating licensing requirements completely, states might at least consider relaxing constraints on those who are licensed as professionals in another state. Although states may prefer to set their own minimum quality standards, and might be reluctant to see them diminished by the lesser norms of other states, they should consider permitting their residents to defer to another state's credentialing as evaluated by an SA search focused on the individual's particular need.

## IV. FACILITATING THE SORTING OF OPTIONS

The burden of compiling and processing data about products often leads shoppers to make choices based on less than full information. SA databases, even better than charts, however, can give consumers the opportunity to retrieve and display lists of products according to any set of attributes that SA firms identify with labels. This product data may be divided into three categories. Part A considers objective, inherent, often quantitative, static characteristics of products and the conceptual categories into which they may fit. Reliable data about these are generally available directly from manufacturers. Part B reviews attributes that measure performance, including evaluations by experts and consumer surveys. Part C discusses data that relate tastes for a product with buyer preferences for other products.

### A. *Objective Inherent Attributes*

The Internet provides consumers and compilers of SAs with easy access to the detailed product information now posted on manufacturers' websites. These include relatively well-defined quantitative measurements of static qualities — such as sizes and ingredients — along with any other relevant categories into which a product fits — such as "100% fat free," "unattached single-family home," or "permanent press." These categories may be defined by individual retailers, trade associations, the media, or interest groups. Where the government finds that private sector action is inadequate, possibly due to anticompetitive strategic actions, it may adopt its own definitions of terms, such as the USDA's definition of "orange juice."

As manufacturers recognize that SAs are creating databases of this information, their desire to increase sales should lead them to provide it to SAs in the format requested. Armed with such data, many sites already permit customers to search for products based on their most significant, objective, basic characteristics. For example, AOL's PersonaLogic.com permits consumers to shop for a camcorder based on its weight, picture resolution, and recording time, and many sites permit buyers to retrieve musical recordings that meet criteria, such as CD, jazz, saxophone, and solo. Collecting data embarrassing to sellers would require more effort, but where the societal benefit from access to such information exceeds the cost of compiling and revealing it, it might be desirable to require manufacturers to disclose such data.

To ensure the accuracy of their databases, SAs might periodically provide each manufacturer with a printout of all the information compiled about its products so that producers could catch unintentional errors or respond to negative information. Moreover, if SAs refused to employ a reasonable mechanism for correcting errors, it might be appropriate for the government to consider such a requirement, along the lines of the Fair Credit Reporting Act, to protect small businesses from dominant firms.

Manufacturers and SAs soon will also likely label products to help buyers find a gift that can be "matched" to a recipient or an occasion being commemorated. For example, if SA databases label musical recordings with the composer's birth date, location of the piece's premiere, etc., buyers will be able to find CDs that not only match a recipient's musical tastes, but also his or her birth date (or at least birth year), hometown, honeymoon location, etc. To increase sales, manufacturers could also label products with any persons, places, or concepts with which they could be connected by some relevant story, and SAs could supplement those lists or simply confirm their accuracy.

Two sub-issues arise with respect to supplier disclosures of basic product information: (1) how to define terms and standards uniformly in a global market, and (2) how to revise existing disclosure rules for a hypertext medium.

Before discussing these issues, it is important to acknowledge that this Article barely touches on policy issues related to so-called "merchant" variables, such as price and availability, although they are often critical to customer purchase decisions. Although the integration of the merchant selection process and the product selection process can and almost certainly will occur, analysis of that combination is left for another day. This Article simply assumes that the SA is using either the data from a real-time merchant search or, more likely, a manufacturer's suggested retail price, in the manner of *Consumer Reports* or ActiveBuyersGuide.com.

1. Defining Terms and Standards in an Increasingly Global Environment

As just noted, it is usually rather easy for SAs to compile objective data about products in a market. Occasionally, however, it is unclear where a new product fits because the definitions of categories are somewhat vague. For example, should a SA characterize a Japanese version of the South Korean pickled cabbage snack as "kimchi," despite its differences with the South Korean version, or should it tag a wristwatch as "water-resistant" based on its Swiss manufacturer's label? Travelers and retailers of imported products have learned to adjust to the different definitions that terms may have in different localities, but as more consumers enter the global arena for the first time, they may be unaware of what problems to expect. Additional action to educate naïve new buyers may be warranted.

Individual producers or an industry's trade association, eager to encourage future sales, should recognize the value of well-defined international terms and standards for each product market and should act to help shoppers find what they are seeking. Thus, when certain online bookstores sought to deceive shopping bots seeking the lowest prices by hiding all of their profit margins in their shipping costs, shopping bots quickly responded by redefining the price used for ranking stores to include shipping and taxes.

Consumer groups or evaluation services may suggest useful standards and definitions, or offer appropriate conversion tables between norms and definitions in different lands or beween different companies' sizes and colors. In some cases, however, competitors' conflicting views may produce stalemates and an industry may be unable to settle on reasonable standards, particularly in a global context. Some firms may seek to exploit the standard-setting process to frustrate competition. In such cases, government involvement may be in the public interest. For example, the Department of Agriculture recently adopted highly praised standards for "organic" food and the courts have enforced violations of the antitrust laws. Where international descriptive standards appeared desirable, industry and international bodies might adopt them. When this proves impractical, government treaties like the one regulating the use of the term "champagne" might be pursued, possibly through the United Nations. Once standards were set, government agencies could police the marketplace for false or misleading product claims.

2. Encouraging Appropriate Disclosure Rules for a Hypertext Medium

In addition to applying existing disclosure rules to cyberspace, the government could also consider whether to take advantage of the Internet to adopt a more market-driven approach to disclosure requirements. Rather than imposing *requirements* — which could have inadvertent negative effects on information flows — the government could focus on helping consumers to *demand* the disclosures that intermediaries, like SAs, Better Business Bureaus, or the Consumer Product Safety Commission ("CPSC") suggested. Manufacturers would face market, not government, pressure to reveal the data sought by intermediaries, including, for example, food ingredients in a searchable form (for those with allergies) or in the seals of approval discussed above, indicating that a respected intermediary was willing to verify some feature or condition. The FTC has already recognized the value of such private-sector certification systems in the context of industry privacy rules, and the CPSC has at times relied on existing industry standards. In addition, existing government educational efforts already promote this type of educated shopping.

*B. Performance Assessments*

Although product brands can often help consumers estimate a product's long-term performance, many shoppers also want to consider expert and fellow shoppers' assessments of a product's performance. At least two issues arise in this area: financing credible product evaluations and the treatment of implicit endorsements.

1. Subjective Evaluations and Testing

   Shoppers often solicit product evaluations from friends, family, and colleagues, but many also seek aid from experts who gather or solicit large numbers of consumer evaluations or conduct their own performance reviews.

   a. Gathering Volunteered Evaluations

   Much information useful to shoppers is produced by third parties pursuing other goals. For example, school test results, high school graduation rates, and crime rate data that home buyers (and thus real estate agents) may want is compiled by school districts and the police, respectively. New York City diners can consult a website listing cleanliness reports compiled by the city's restaurant inspectors. There are many ongoing efforts to assess the quality of health care plans according to statistically valid, quantitative data, and public health entities collect data on medical success or error rates as well as adverse actions against physicians. In addition, some social activist groups provide seals of approval for companies that meet their standards.
   SAs already store data about products that have won or attained finalist status for prestigious awards, with some book sites maintaining data about more than 50 different prizes. SAs might also quantify the wealth of quality assessments available from a variety of sources. These include negative comments, such as medical errors found by peer review groups, complaints filed with Better Business Bureaus and state consumer protection offices or at websites created by disgruntled customers, as well as both negative and positive comments posted at discussion groups like alt.fashion, at manufacturers' online communities, and at shopping sites like Amazon. Repair records could also be compiled more easily if uniform survey standards could be developed. SAs that could gather all this data, while screening out propaganda, would provide useful assistance to buyers. Ideally, an SA could quantify these overall quality assessments according to a numerical scale to help consumers better sort products according to these assessments.

   b. Financing Expert Evaluations and Consumer Surveys

   Shoppers also value large, random samples of consumer reviews, and thus they buy Zagat's restaurant guides. Similarly, manufacturers finance surveys by respected firms like J.D. Power and ValueStar.com. Then there are newcomers spawned by the Internet, like Epinions.com, deja.com, and others that use multi-dimensional ratings. These reviews and rating systems, however, are susceptible to unconscious biases, rings of conspirators, and other factors that diminish their credibility.
   Therefore, buyers place even greater value on the evaluations of respected experts who test or grade products themselves, such as *Good Housekeeping* and *Consumer Reports.* In addition to these institutions, consumers also rely on individual reviewers, other independent experts, and "mavens." Because their credibility is essential to their role, these advisors attempt to resist pressure from product producers. Moreover, as wireless communications devices become more prevalent, these experts may encourage instant customer feedback confirming their assessments or suggesting that a reevaluation may be warranted.
   SAs, like ConsumerSearch.com, can collect all scores published by value-rating services for relevant product markets. Manufacturers also have a strong incentive to provide SAs with all positive reviews of their products and perhaps even negative reviews about their competitors' offerings. SAs with special expertise in the product market may also add their own assessments to their databases, as real estate agents often do. Yet some observers have expressed concerns that consumer product evaluation services would be undersupplied in the market if product evaluators were unable to control the dissemination of their results.

As discussed above, the combination of databases and the Internet dramatically amplifies the potential for free riding and may justify the additional protection offered by H.R. 1858 against mere copying. Although one can copyright subjective product evaluations, including even estimated fair market prices, laws that prohibit them from being used as inputs to a private, searchable database would appear to suffer the same infirmities as H.R. 354. Meanwhile, it appears that the marketplace already offers firms sufficient options for responding to this free rider problem.

First, entities like Consumers Union could seek payment from other interested parties, including the firms whose products they were evaluating. Although Consumers Union might reject this latter option for fear of damaging its credibility, it should be noted that auditors do not lose their credibility when businesses pay them to verify financial records or privacy policies. It also seems clear that public broadcasters, as exemplified by *The NewsHour with Jim Lehrer*, are able to maintain their independence despite acceptance of grants or brand commercials from parties on whom they report. Although there is certainly a credibility problem for entities that report on their owners or face other conflicts of interest, the key to keeping their reputations intact appears to be careful disclosure of any conflicts that others may perceive.

Second, evaluators could refrain from publishing their empirical data. They could offer their customers the chance to sort among choices based on the raw data, but without actually seeing it displayed, thereby relying on trade secret law to protect valued intellectual property. Consumers could search for products that were rated above average in a few specific categories or rated highest in one category and above average in a few others, or any of a myriad of other combinations. Testers could also license the use of their raw data to others who agreed not to display it. As another option, product testers could provide or oversee testing for others, as *Good Housekeeping* agreed to do for the website brandwise.com.

If neither of these incentives were sufficient, foundations might step in to fund the collection of such data. Academic research centers could also move into the breach to fund independent evaluations, like the annual Airline Service Quality studies. Alternatively, government funding might be justified where a cost-benefit analysis indicated that the research was worthy, but the "public good" aspects of the information, i.e., that it would be available freely to everyone, would discourage the private sector from pursuing such studies. Research on auto safety, toy safety, airline on-time performance, or regarding food and medical practices would likely fall into this category. The Internet would also increase the justification for government funding of product testing because information available online would be more accessible to the public.

c. Specialization and Professional Certifications

Because some buyers may care solely about the product's performance at one or two tasks, rather than its overall rating, compilers of customer ratings generally grade different aspects of a product or service separately. For example, Zagat's and deja.com provide separate ratings for a restaurant's food, atmosphere, and service, and *CR* rates many different relevant elements of product performance in its charts. Similarly, many respected retailers and magazines focus on one aspect of a product, e.g., "cutting edge" fashion or top-of-the-line performance for serious cyclists or a specific niche of books. Not only could SAs compile all of these specialized ratings, but they could also translate the comments previous customers made in forums — on- or off-line — into separate and distinct ratings for different uses.

In fact, most consumers seeking services from a professional are only concerned with the supplier's performance for that specialized task, rather than their skills outside that specialty. SAs could serve consumers well by enabling them to focus on specific performance qualities and documented expertise, while ignoring competency in less relevant areas. For example, one company has compiled a database of 400,000 people to help students find the best college for attaining the specific lifestyle they seek. Consumers could use the customized search profiles discussed below to sort product or service options based on any combination of different attributes.

2. Reporting Implicit Endorsements

Consumers also rely on the implicit endorsements of respected stores, magazines, or salespeople that sell, display, or use products without any explicit comment. In fact, consumers may often rely more on these retailer "brandings" than manufacturer brands. SAs that label products to show such implicit recommendations could give consumers the confidence to consider new and lesser-known manufacturers' products by labeling them with "icon" brands. For example, SAs could label a restaurant as having a "Brooks Brothers" style or jewelry as having a "Wired Magazine" style. SAs might also label products with the names of well-known and respected buyers that shoppers might want to imitate, for example, the linens used in the Waldorf Astoria hotel and the desk chairs purchased by Goldman Sachs. In addition, SAs could include the choices made by celebrities, fictional characters, or anyone else that customers might want to emulate. For example, SAs could supplement labels to help shoppers retrieve the lipstick worn by Monica Lewinsky on her 1999 20/20 interview with Barbara Walters and outfits worn by a favorite character on a TV show. Irrespective of the objections of celebrities, the First Amendment appears to give SAs free reign to use facts about the products celebrities have displayed in public.

The value that consumers place on a product may also depend not only on who else uses it, but upon the number of other purchasers. Some may prefer a product, like a "Harry Potter" book or a book featured on *Oprah*, which they can discuss with many others. Others may prefer more exclusive goods. To enable consumers to search by this attribute, SAs could add product sales data to their databases.

For buyers seeking a gift, the most relevant evaluation is that of the recipient, and thus many retailers have expanded their bridal registries to "any occasion" lists; meanwhile, companies like WishList.com have established online versions. Wives or husbands can aid their spouses with gift ideas, and retailers and SAs alike might consider mining these sites for their data.

Current case law would appear to protect the right of SAs to collect and use data about consumer shopping preferences, such as these online wish lists. If customers with gift registries desired greater privacy, then SAs could easily limit access by requiring a password to view gift data. Meanwhile, upscale stores could try to respond to free riding on their expert quality assessments in at least two ways. First, they might attempt to shift further toward private label products. Second, they could emphasize their comparative advantage in helping customers find the most suitable products by designing superior search profiles, as discussed in Section V below.

### C. Collaborative Filtering Data: Implicit Features

In some cases, shoppers browsing for a product, like a book or CD, may be unable to articulate some of the key attributes of what they are seeking, particularly the taste-based "experience" qualities discussed above. The relatively new and powerful process of collaborative filtering ("CF"), however, is now permitting consumers to search for products based on the implicit recommendations of others who appear to have very similar tastes. CF software solicits user preferences implicitly by asking users to rate previous and subsequent purchases. Then it identifies all other users in its database who exhibit similar preferences. From this "similar tastes" group, it can retrieve ratings of a specified product or identify the item rated highest by the group. These recommendations appear more likely to predict what consumers will like than even the expert critics whose tastes often differ from consumers'. Similar systems might help consumers identify new products most likely to complement products they already own.

CF has become more practical as technology has diminished the cost and difficulty of collecting and processing consumer transactional data and rewarding respondents enough to overcome their general apathy. Moreover, CF database compilers are finding it more practical to correct the error-prone presumptions of primitive algorithms — that all of a consumer's purchases represent his or her preferred choices — when consumers are disappointed with purchases or when they have purchased items for others. CF compilers are also starting to help customers benefit from the use of both explicit and implicit data. As developers experiment with new varieties of CF systems and CF becomes increasingly more effective, network effects may create a danger of monopolization and sabotage could also arise, as discussed below.

1. Network Effects and Winner-Take-All Markets

Whether CF databases will become natural monopolies depends on many factors. All other things being equal, a CF database with more entries should produce better results than a smaller one because it is more likely to include closer matches for any given customer. The differences may be insignificant, however, when the smaller database includes a large enough number of individuals to produce excellent results. Still, small differences may matter for major purchases like a two-week vacation package. Furthermore, as customers seek to match increasing numbers of search criteria, the largest CF databases are likely to produce significantly more accurate results.

Two factors suggest that consumers will seek matches on more criteria. First, because consumer tastes in different product markets are often related, CF databases should be more effective when they combine data from multiple, potentially related product markets, including restaurants, films, vacation spots, etc. Moreover, to the extent that such correlations are substantial but privacy rules prohibit firms from sharing data with other SAs, mergers and consolidation might be undertaken to facilitate the creation of such multi-market transactional databases.

This opportunity for aggregating such data undoubtedly motivated Amazon to offer to host other retailers on its site and handle their transactions. The infomediary firms proposed by Hagel and Singer would achieve similar results on the customer's side. Consumers' satisfaction with CF systems might well override any of their privacy concerns about data aggregation. Second, because consumers' preferences are often significantly affected by their most recent purchases, consumers might seek recommendations from only those who had made similar recent purchases.

Many factors may also work against the emergence of natural monopolies. Diseconomies of scale and scope may make it impractical for a single entity to focus its resources on evaluating too many different facets of each product choice. If so, specialized databases might emerge to focus on all of the variables most important to a target segment of customers, offering them superior quality specialized service. Network effects — and the creation of natural monopolies — may also be limited by consumers' desire to compartmentalize different aspects of their purchasing decisions among multiple SAs to protect their privacy.

If quasi-natural monopolies do arise, antitrust laws will require policymakers to prevent firms with market power from bundling their services to distort competition in related markets. These concerns about bundling services highlight the need for ways to distinguish among separate product or service markets and to identify arrangements that represent illegal tying. Although it might not be in the public interest to treat the information in such databases as essential facilities for other businesses, this issue may deserve further analysis.

2. Sabotage of Collaborative Filtering Databases

As CF data becomes more and more useful to buyers and sellers, SAs may well compete intensely based on the quality of their CF data and predictive algorithms. Although there is no way to ensure that all consumer entries included in a CF database are accurate, deterring the intentional submission of false data would clearly be in the public interest. Current tort laws may already forbid such submissions as tortious interference with business interests, but it would also probably be helpful for government bodies to help businesses spread the word that the "pollution" of CF databases, by intentionally submitting false data, is prohibited by law.

*D. Managing Congestion*

Whether or not eBay was likely to face congestion problems due to Bidder's Edge and others, congestion on the Internet is not a trivial problem. Moreover, it will probably get worse if consumers increasingly rely on shopping bots to make Internet price comparisons, leading merchants to employ "pricebots" to adjust their prices based on the continuous monitoring of their competitors. If retailers felt forced to add additional capacity to ensure that bona fide customers were not blocked from their sites, they would have a strong case for recovering their additional capacity expenses from those parties causing the unwanted congestion.

It is not clear that tort law provides an efficient way for retailers to prevent such congestion. In July 1996, a Best Buy store in Reston, Virginia had police remove an individual who was copying down prices of its television sets, but a Fairfax judge quickly dismissed trespassing charges without even considering whether the customer's behavior had been disruptive. Some earlier cases, however, have gone the other way. Nuisance laws might discourage congestive practices, but leaving these issues to ad hoc decisions by judges does not appear to be the most efficient alternative.

Ideally, technology will solve the problem. A solution may arise in response to the comparable problem created by "spiders" and other web crawlers used by search engines to index the web. In the interim, firms may adopt new strategies for managing the problem. EBay already permits aggregators to search its website during off-peak hours and to do searches "on demand" for buyers. Firms might also create alternative URLs or listservers to provide price changes or other information in an efficient manner for a fee.

Alternatively, trade associations and government agencies could try to formulate guidelines to minimize congestion by setting a level of frequency of pricebot requests that would represent prima facie evidence of a nuisance. Retailers who offered access to their freely available data on reasonable terms should then be permitted to recover damages for harassment or nuisance from those who congested their sites beyond the designated levels.

## V. DEFINING A BUYER'S SEARCH PROFILE

Some consumers already know the combination of attributes they are seeking in their desired product. Many others, however, are likely to desire assistance when trying to compile an effective search profile, i.e., the characteristics on which to focus and how much weight to give to each. Most consumers have learned to adopt strategies and construct preferences to make their searches manageable, if somewhat less accurate, rather than bucking the limits of their information processing abilities to pursue their optimal choice.

One difficulty in designing precise search profiles is that it requires consumers to be clear about what they want, including all of the conditions, goals, and aversions relevant to their purchasing decisions. Then they must translate these into precisely quantified preferences for particular product features. Given historical constraints on information processing, the incremental benefits from such efforts are unclear, and consumers are rarely tempted to undertake such a project.

SAs, however, can substantially improve the accuracy of consumer decision-making in two ways. First, computers have the information processing power to use a profile to evaluate each and every option. Second, SAs can offer consumers new and easier ways to construct optimal search profiles that (1) eliminate all selections that fail to meet their minimum requirements, and (2) rank acceptable options based on the intensity of shoppers' preferences for relevant attributes. Section A discusses the pursuit of these two goals in the absence of SA assistance, and Section B discusses the new options that SAs can offer. Section C analyzes some potential dangers to competition in this area.

### A. Identifying Existing Conditions and Preferences

1. Absolute Requirements

Although consumers generally "want it all," most are willing to accept items that lack some desired features or include some undesirable characteristics. In fact, most shoppers expect to make tradeoffs among their multiple preferences for product attributes. Consumers, however, usually have some absolute requirements, i.e., "non-compensable" preferences. These may include budgetary limits, timing deadlines, or spatial and geographic restrictions. Additional constraints may arise from strong consumer tastes, like an insistence on wearing only black clothes or an abhorrence of country and western music.

Consumers are usually aware of most of these absolute requirements before they consider a purchase, but they may overlook or be unable to ascertain others, such as whether a new purchase, like piece of computer software, is compatible with their relevant existing possessions. In addition, shoppers may lack the expertise to recognize the significance of some conditions. For example, a novice cyclist may be unaware of what features to demand or refuse when purchasing a bicycle for a long trip through unusual terrain.

2. Compensatory Preferences: Goals and Conditions

Selecting and weighting compensatory preferences for ranking a buyer's acceptable choices is a more difficult task. To do this with precision, consumers would, at least in theory, need to ascertain their "utility function" and translate it into a formula for evaluating each of their options in a product market. Compiling their utility function would generally require them to articulate both their fundamental and derivative goals and the appropriate weights for each. Constructing the latter product-specific formula would seem to require them to (a) identify each of the attributes of the product that they find significant, (b) attach a subjective value to each possible variation of that attribute, and (c) assign a weight to each feature based on its relative importance.

This project would represent a impractical, if not impossible task. Moreover, the inherent subjectivity of this process is apparent from the controversy generated when magazines try to rate the best colleges or cities. Finally, additional complexity arises because many consumers find variety and change pleasing, if not productive. Therefore, their preferences for many products are often significantly affected by their most recent purchases in the product area.

### B. *Evaluating, Modifying, Anticipating, and Applying Preferences*

Under these circumstances, even sophisticated shoppers may want assistance with the tasks just discussed, particularly with measuring conditions, assessing the consistency and consequences of their preferences, and understanding the details of a product market. SAs can help consumers by making it much easier for them to use much more accurate ranking systems, like weighted adding.

1. Prompts, Examinations, and Suggested Defaults

Most shoppers are likely to benefit from prompts, which anticipate responses with default answers. For example, doctors, real estate brokers, and investment advisors often give new clients checklists of possible conditions and objectives to jog their memories for pertinent information. Moreover, improvements in voice recognition technologies should soon allow these questions to be asked and answered orally, even by telephone, from remote locations and at convenient times. Alternatively, animated electronic agents may be able to handle this task, particularly if they can make the small talk that builds customer trust. In addition, expert systems will be able to process answers to initial questions quickly enough to allow them to offer anticipated default answers to later questions and automatically edit out inapplicable ones. Moreover, questioning could reveal consumer preferences implicitly by asking consumers to choose which of two options was better, in much the way that eye doctors establish a patient's optimal lens features. Many manufacturers' websites already offer "configurators" to help consumers create search profiles for seeking the best product matches on the website.

In many cases, consumers may also want an expert to observe and assess their condition. For example, they might desire a beauty consultant or tennis pro to examine their strengths and weaknesses and recommend product attributes best suited for their needs. Experts could assist online shoppers by conducting remote examinations or by supervising self-examinations by consumers. For example, IC3D.com asks customers to measure the eleven specific body lengths it uses to produce customized jeans. To facilitate expert assessments, consumers may also submit photos, create virtual models, or even undergo holographic body scans.

As indicated above, probably the most difficult task facing consumers is quantifying the relative weights of different preferences. Fortunately, new technologies can provide helpful tools. For example, to help federal government employees identify the best health care plan for their needs, the Office of Personnel Management offers them use of PlanSmartChoice software, which, in 2000, worked in four steps. First, users were asked to select from a list the features they considered most important. Next, they were asked to consider how important it was to find a choice that was rated highly with respect to each of those attributes. Third, users were asked to compare pairs of offerings and to quantify how strongly they favored one option over the other. PlanSmartChoice software then used this consumer input to estimate weightings, generate ratings, and rank the available plans.

A more promising alternative for managing this task is the opportunity that some SAs give consumers to select a profile based on a few descriptive terms about themselves. For example, consumers shopping for a camera using software developed by companies like frictionless.com can select the profile for a "budget buyer," "beginner," or "expert," and use the set of default preference weightings suggested by the SA for each. In fact, SAs can construct profiles to reflect every type of buyer imaginable, beginning with those represented by the hundreds of specialized magazines and boutiques, i.e. brands, with which shoppers might identify. By incorporating demographic and lifestyle data, SAs can help consumers identify the best default profiles, beginning with those based on their "VALS" lifestyles.

Furthermore, while both editors and retailers generally, at least implicitly, use profiles of their target audience to produce only a single version of their presentation, a database medium permits SAs to offer customers the opportunity to customize a version of that profile before applying it to the data set. That is, consumers could tinker with the default formula and modify it to reflect their own idiosyncratic tastes. Many online publishers already permit viewers to design profiles to personalize displays on their web pages. Databases could permit a woman shopping for evening wear to customize her standard Bloomingdales profile to favor particular fabrics or a specific fashion editor's tastes. Similarly, a student could modify the profile used by a college guide to give different weights to specific academic departments or sports programs. Consumers could give positive or negative weight to any features that SAs labeled in their databases, including particular expert credentials. Since a manufacturer's brand generally represents a specific set of features and qualities, consumers could also be offered a chance to select variations on or combinations of the brands they liked. Finally, SAs could offer shoppers a third option — the opportunity to select a profile implicitly by using CF to generate recommendations, as discussed above.

2. Reviewing and Evaluating Preferences and Conditions

Clever marketing practices that lead consumers to employ misleading theories may cause buyers to adopt preferences that are inconsistent with their long-term goals or conditions. Consumers may also confuse novelty with longer-term value, misestimate their needs, or fail to give appropriate weight to relevant factors, or be induced to bias their decision making structure.

To help them avoid such unwanted consequences, shoppers may seek consultants to review and interpret their purchase histories. SAs might offer these services themselves or refer consumers to experts, such as those listed on sites like ExpertCentral.com and EXP.com. These consultants could point out which of a consumer's constraints or conditions were unnecessarily eliminating potentially desirable choices and might easily be relaxed. Consultants might also help consumers reconcile conflicting goals or recommend training courses — for gaining the prerequisite skills needed for some options — or psychotherapy, to eliminate destructive preferences. Consultants might also be helpful to those inclined to rely on CF, but who wanted to shed undesirable preferences reflected in their past choices. Since consumers are generally reluctant to reveal data that might embarrass them, these consultants would first need to earn their trust.

3. Information About Features and Trends that Might Alter Preferences

Information about trends and predictions about future developments are also likely to affect shopper preferences. Just as car buyers adjust their purchase decisions to account for the annual introduction of new car models, consumers might want to refrain from buying a type of computer system in December if a new "breakthrough" version or a significant price cut was expected in March. Consumers might also want to consider the financial health of a product or service provider when they expect to have future contacts with the provider and the industry is volatile.

In some cases, consumers might desire further information about a particular feature. In response, many comparative shopping services permit consumers, with a simple click, to display detailed explanations of the feature or capability. Some offer software programs, such as Broderbund's Cosmopolitan Fashion Makeover, to permit consumers to experiment with offerings, while other online retailers, like Lands' End and FurnitureFind.com, offer access to live salespeople. Still others have recognized the advantage of "brick-and-click" combinations with conventional stores.

Shoppers can also refine their search profiles by browsing for ideas about what features to favor or avoid. Browsing can prompt them to remember forgotten or unconscious preferences or to learn about new features. For some consumers, browsing among pleasing options can be comparable to sightseeing or visiting a museum. For others, SAs and retailers can offer more entertaining forums, including customized versions of interactive games, supported by ads or user fees. Furthermore, once consumers have adopted a profile and reviewed the recommended items, they may want to alter their profiles to readjust the weights of some features. This process could go through multiple iterations.

Websites, such as ConsumerSearch.com, also include more general background information and direct shoppers to reviews of a product category. Many include postings of frequently asked questions ("FAQs") or forums for querying fellow shoppers. Trade publications may offer more expertise, but they may also be biased due to strong pressures from advertisers. Consumers could also consult independent entities, like *Consumer Reports* or a trusted author or government official. These sources might also offer customers candid assessments of different expert rating services, warning consumers about services with credibility problems, or recommending those likely to use a target profile most like the buyer's.

While retailers might try to offer free consultations as a marketing tactic, the increased use of shopping bots could create conflict between the cost of providing expert consultations and the need to maintain competitive prices. While technology may help cut the telecommunications expenses of such assistance, the costs of providing expertise may force retailers to unbundle and charge separately for them, as doctors and cosmetics retailers often do.

### C. Eliminating Stickiness Due to Ignorance

The potential for higher profits encourages firms to improve the attractiveness of their products or services so as to increase the likelihood that customers will remain with or repeatedly return to them — known as "stickiness." It does not follow, however, that firms should be rewarded for the higher "switching costs" that seem likely to result from consumers' current failure to appreciate the value of their purchase history data. Rather, it is probably desirable for the government to educate consumers about the value of maintaining control over their transactional data as well as techniques for shopping online most effectively, thereby minimizing their switching costs.

1. Granting Consumers Control Over Their Data

As discussed above, data about past purchases is often crucial to enabling shoppers to search most effectively, particularly with respect to collaborative filtering. Nevertheless, consumers often forget some of their explicit preferences or fail to discover implicit ones. With computers and databases making it easier to track purchases passively, many firms now compile such data so that they may use it with CF to anticipate consumer choices and make recommendations. In fact, firms with multiple retailing channels are beginning to synchronize their catalog, store, and website sales records, as well as other consumer data. Firms are realizing that an excellent way to create stickiness is to get customers to leave something of themselves behind.

Yet if consumers become enamored with the effectiveness of CF and refuse to give up its full benefits, this may create a significant entry barrier for competitors, even those that might offer superior CF algorithms. While firms will presumably be willing to sell each customer his or her data at some price, there would seem to be little incentive for those firms to limit their requests to cost-justified prices.

Now consumers need not rely on their former CF providers to maintain that transactional data. Instead, they can use software products and infomediaries to collect it. Meanwhile, competition between these groups should ensure that such data collection services are available at cost-based prices. Yet this presumes full information, while most consumers today are unaware of these alternative data collection mechanisms. More importantly, most shoppers are not yet familiar with CF, and even those who are may not fully appreciate how effectively CF can help them to avoid bad choices. Shoppers' failure to foresee this value and respond appropriately could lead current firms to gain an insurmountable early advantage over new entrants with better offerings.

Nevertheless, many support a hands-off policy — permitting "first movers" to exploit any monopoly they enjoy over data they collect — in the belief that increasing their reward will stimulate initial innovation. On the other hand, permitting excessive advantages to accrue to the first innovator might also hinder subsequent inventiveness in this area. The European Union has taken the opposite position, requiring all businesses collecting transactional data about a customer to provide that data to the customer free of charge. Yet absent the opportunity to earn a return commensurate with the riskiness of their investments, data collection entrepreneurs might well forgo investing in technologies to improve data collection.

One possible model for resolving the issue would be to follow the approach that American law gives to patient medical records. Although the law generally gives ownership of that data to doctors, more than forty states have given patients either a statutory or common law right to some type of access to their medical records, and many states limit the price that patients may be charged for copies of those records.

Alternatively, to encourage the marketplace, rather than regulators, to set any prices or specific procedures, public officials might simply use speeches and other public appearances and writings to inform consumers of the value of their data and to recommend that they demand disclosure of these terms and consider alternatives for maintaining copies of their own data. The government could include such advice in public service announcements and school lessons, and public agencies might encourage trade associations to establish standards or to develop technologies to facilitate the compatibility of data formats used by different firms. Policy makers might also consider a temporary disclosure requirement to give customers a frame of reference on this issue. If this policy was successful, consumers would demand such disclosures before they chose a retailer, or at least before their existing retailer had collected too much data about them. Then market competition would lead retailers to offer shoppers their data at cost-based prices, such as a flat rate or one based on the time period involved or the quantity of data collected.

2. Teaching Consumers to Use New Search Tools

Historically, children learned shopping skills by observing family and friends and by studying arithmetic in school. Today, however, with the emergence of databases and the explosion in the number of product choices, the long-standing shopping skills of old are no longer sufficient in the new Internet environment. Retailers that help consumers become familiar with navigating their websites will likely create significant switching costs beneficial to the firms, but potentially harmful to consumers. Therefore, public schools and adult education centers should consider teaching students how to use these new tools most effectively. Libraries might also offer consumer education programs, possibly taught by volunteers from the community.

The FTC has already published numerous electronic consumer and business guides, and with the help of librarians and others — providing summaries of the best materials as well as links to them — consumers could learn to become educated, demanding, and effective shoppers. Educators could also teach consumers about the long-term benefits of embracing flexible, "open access" search tools, which work on nearly all websites.

3. Unbundling Data Identification and Consultation

As discussed above, many traditional retailers help consumers compile search profiles free of charge in return for expected purchases. This makes sense when such assessments would be difficult to sell as separate services and where consumers are relatively captive. When there is consumer demand for assistance in compiling a search profile as a separate service, such as obtaining holographic body images to help customers shop more effectively for clothes online, one would expect the marketplace to provide it. If firms attempted to use bundling to extend market power they held in one service market segment into another, then the current prohibition against tying should be recognized as requiring unbundling to foster competition.

*D. Incentives for Developing Search Profile Templates*

While SAs have the option of using existing trade secret laws to protect their proprietary database entries, that option would not be available to SAs that offered shoppers the opportunity to customize specially designed search profile templates; those SAs would need to reveal their templates for consumers to use them. Fortunately, it appears that existing copyright law would protect SAs who develop creative and valuable search profiles. Although the attributes they use would not be copyrightable, the pattern of values and weights they employed would presumably represent a novel variation of a search profile. While some entrepreneurs in this area may have already sought patent protection for such profiles, it would seem inappropriate to grant any firm a business method patent on a system that is already used implicitly by every business with a target audience in mind. Firms with reliable reputations that designed valuable search profile templates could license them to others, but new entrants could still design their own variations and improve upon the existing templates as products and consumer tastes evolved.

VI. EXAMINING INDIVIDUAL OPTIONS IN DETAIL: FREE RIDING AND RETAILER DISPLAYS

Once shoppers have used a profile to search all options and identify their most suitable choices, they are likely to want to examine those items in more detail. Furthermore, because the adoption of a profile can be so imprecise, consumers may wish to view their choices and then readjust their profiles several times before making a final decision.

Yet, it is not clear what role SAs will play as providers of product displays, given the increased potential for free riding. As online consumers are beginning to discover, a click or two via a shopping bot like Deal Time, R U Sure, or any of the many listed at BotSpot.com can transport them from the best online product displays to a trusted low-cost merchant. Some, like dash.com, can even launch an automatic search for competitor prices whenever the user views a product on a shopping site. While one 1999 study found no evidence of significant free riding on the Internet, that was probably because it takes time for consumers to become familiar with reliable shopping bots. Although some online retailers have sought to block access by shopping bots, such efforts have not defeated the bots. Moreover, most of the deterrents that protect against free riding in the context of brick-and-mortar stores have dramatically weaker effects with respect to online shopping.

Free riding on the Internet seems almost certain to increase as consumers find it easier to exchange referrals to low-cost sellers, particularly for the many shoppers attracted to the Internet primarily to obtain low prices. E-mail already makes it simpler for buyers to use what has been called "word of mouse" or "viral marketing." If friends send a properly formatted website address to those on their e-mail lists, recipients need only click on the address to reach the recommended website. Some consumers may spread the word for altruistic reasons, and retailers can encourage recommendations by using the Amway formula of rewarding customers who recruit new users. Communication of this information will be even easier once voice recognition technology software makes Internet communication — and shopping bot use — as consumer friendly as voice telephony.

Retailers who create online product displays may respond to free riding by asking manufacturers to reimburse their costs, as REI already has, but manufacturers are likely to refuse. Instead, manufacturers will probably find it more cost effective to finance a single, high quality display of each item on their own websites, available to all online retailer customers via hyperlinks. These webpages might include 3D displays, like those already available at The Sharper Image, other high-tech offerings, or even infomercials on demand.

If shoppers in traditional stores begin to use portable PCs to consult shopping bots for lower prices, free riding at these stores could increase substantially. This could lead manufacturers to agree to compensate traditional retailers for their physical displays, then publicize retailer addresses to online customers who want to make physical inspections. Manufacturers could even pay for traveling displays, in which they bring products to potential customers' homes for inspection.

## VII. CONCLUSION

Comparison shoppers, like all decisionmakers, seek two types of information: a short list of their best options and detailed information about those choices. The Internet now gives them direct access to most of the detailed descriptions they need to make purchasing decisions, but many consumers still seek assistance selecting the most suitable products. Selection assistants now have the technical capabilities to provide consumers with near-optimal service by (1) compiling comprehensive, searchable databases with all of the attributes by which shoppers may want to search; and (2) helping buyers effectively search these databases using carefully crafted search profiles. These services have the potential to improve the efficiency of consumer product markets, benefitting consumers and the most valuable producers alike. Policymakers should try to maximize the benefits of market competition in this area by eliminating obstacles to competition not likely to be overcome by other means.